\documentclass[aps,showpacs,twocolumn,eqsecnum,floatfix]{revtex4}
\usepackage{latexsym}
\usepackage{epsfig}
\usepackage{amsmath}
\usepackage{mathrsfs}

\begin{document}

%%%%%%%%%%%%%%%%%
%%%   TITLE   %%%
%%%%%%%%%%%%%%%%%

\title{Regularization of spherical and axisymmetric evolution codes in
numerical relativity}

\author{Milton Ruiz}
\email{ruizm@nucleares.unam.mx}

\author{Miguel Alcubierre}
\email{malcubi@nucleares.unam.mx}

\author{Dar\'\i o N\'u\~nez}
\email{nunez@nucleares.unam.mx}

\affiliation{Instituto de Ciencias Nucleares, Universidad Nacional
Aut\'onoma de M\'exico, A.P. 70-543, M\'exico D.F. 04510, M\'exico.}

%%%%%%%%%%%%%%%%
%%%   DATE   %%%
%%%%%%%%%%%%%%%%

\date{\today}

%%%%%%%%%%%%%%%%%%%%
%%%   ABSTRACT   %%%
%%%%%%%%%%%%%%%%%%%%

\begin{abstract}
Several interesting astrophysical phenomena are symmetric with respect
to the rotation axis, like the head-on collision of compact bodies,
the collapse and/or accretion of fields with a large variety of
geometries, or some forms of gravitational waves.  Most current
numerical relativity codes, however, cannot take advantage of these
symmetries due to the fact that singularities in the adapted
coordinates, either at the origin or at the axis of symmetry, rapidly
cause the simulation to crash.  Because of this regularity problem it
has become common practice to use full-blown Cartesian
three-dimensional codes to simulate axi-symmetric systems.  In this
work we follow a recent idea of Rinne and Stewart and present a simple
procedure to regularize the equations both in spherical and
axi-symmetric spaces. We explicitly show the regularity of the
evolution equations, describe the corresponding numerical code, and
present several examples clearly showing the regularity of our
evolutions.
\end{abstract}

%%%%%%%%%%%%%%%%
%%%   PACS   %%%
%%%%%%%%%%%%%%%%


%%%%%%%%%%%%%%%%%%%%%
%%%   MAKETITLE   %%%
%%%%%%%%%%%%%%%%%%%%%

\maketitle

%%%%%%%%%%%%%%%%%%%%%%%%
%%%   INTRODUCTION   %%%
%%%%%%%%%%%%%%%%%%%%%%%%

\section{Introduction}
\label{sec:introduction}

After forty years of research, the black hole collision problem can
finally be considered solved. Though there are certainly still many
details to be worked out, the results from
Pretorius~\cite{Pretorius:2005gq}, the Brownsville and the Goddard
groups~\cite{Campanelli:2005dd,Baker:2005vv}, and other groups that
have followed, show that it is now possible to follow the numerical
evolution of two black holes for several orbits, through the merger
and subsequent ringing of the final merged black hole.

Such tremendous progress in full three-dimensional (3D) numerical
relativity, however, does not imply that there are no more interesting
astrophysical situations that can be studied in spherical or axial
symmetry, such as for example gravitational collapse or accretion onto
compact objects.  Accurate 3D simulations still require large
computational resources, so that exploiting the existing symmetries
should allow important savings in computational time. Using
coordinates adapted to the symmetry, the number and complexity of the
evolution equations are reduced and thus the computational cost is
also reduced.

Nevertheless, the development of general purpose
spherical/axi-symmetric codes in numerical relativity has been
hampered by the lack of a generic method to deal with the
singularities associated with the symmetry-adapted coordinate
systems. For example, in the spherically symmetric case described with
spherical coordinates $\left(r,\theta,\phi\right)$, the coordinates
become singular at the origin $r=0$. This implies that several terms
in the evolution equations diverge as $1/r$, and even though local
flatness guarantees that analytically all those terms should cancel,
such exact cancellation usually fails to hold in the numerical
description. A similar problem arises in systems with axial symmetry
when approaching the axis of symmetry.

Several methods to deal with this problem have been proposed in the
past. For example, one can choose specific gauges that either
eliminate or ameliorate the regularity problem such as the {\em areal}
(or {\em radial}) gauge in spherical symmetry, where the radial
coordinate $r$ is chosen in such a way that the proper area of spheres
of constant $r$ is always $4 \pi r^2$.  Similarly, in axial symmetry
one can use the shift vector to guarantee that some metric components
always vanish thus reducing the problem of regularity at the axis (for
details see {\em e.g.}~\cite{Bardeen83,Abrahams:1993wa,Evans:1985hg}).
Furthermore, there has been a lot of work on the construction of axial
codes that ensure that the metric remains smooth on the axis.  For
example, Garfinkle and Duncan describe in~\cite{Garfinkle:2000hd} a
method that consists on the introduction of auxiliary variables which
allow one to impose all the required regularity conditions on the
extrinsic curvature. However, this method requires to solve, on every
time slice, an elliptic equation for the lapse, the shift components
and the conformal factor. A similar algorithm was presented by
Choptuik {\em et al.} in~\cite{Choptuik:2003as}, but adapted to the
$(2+1)+1$ formulation.  Recently, another regularization procedure was
described in detail for the Z4 system~\cite{Bona:2003fj,Bona:2003qn},
by Rinne and Stewart in~\cite{Rinne:2005sk,Rinne:2005df}, again also
adapted to the $(2+1)+1$ formulation. A different idea is the
so-called ``Cartoon method'', which consists in evolving three
adjacent planes in Cartesian coordinates and then performing a tensor
rotation to obtain boundary conditions~\cite{Alcubierre99a}. However,
as this method uses a tridimensional code, it is still more
computationally more expensive than an axial code (and requires one to
write a full 3D code in the first place).  We believe that there is
still a need for a code able to keep the equations regular in
curvilinear coordinates while still allowing quite general gauge
choices.

Recently, one of us~\cite{Alcubierre04a} presented a general procedure
to deal with the irregularities at the origin in the case of spherical
symmetry. Such procedure essentially consists in the introduction of
auxiliary variables which allow one to impose all the required
regularity conditions on the metric coefficients. This method,
however, cannot be easily extended to the case of axial symmetry
without spoiling the hyperbolicity of the system evolution equations.

In this paper we follow the idea presented
in~\cite{Rinne:2005sk,Rinne:2005df}, and we use the general form of
the tensor components in an axially symmetric spacetime to show that
one can develop a generic algorithm for regularizing the evolution
equations in both axial and spherical symmetry.  We start by writing
the general form of the spatial metric with the corresponding
symmetry.  After that, we analyze the different conditions that the
geometric variables must satisfy at the origin or the axis of
symmetry.  These conditions arise both from parity considerations and
local flatness.  We then introduce new variables as combinations of
metric components whose parity properties guarantee that both types of
conditions are satisfied at the same time, and evolve those variables
instead of the original metric components.

This paper is organized as follows.  In Sec.~\ref{sec:evolution} we
present the dynamical variables and the evolution equations, both for
the ADM system and for a strongly hyperbolic formulation.  In
Sec.~\ref{sec:spherical} we introduce the regularization procedure for
the particular case of spherical symmetry.  Later, in
Sec.~\ref{sec:axial} we generalize this regularization procedure for
the case of axi-symmetric spaces.  In Sec.~\ref{sec:examples} we show
some numerical examples in both spherical and axial symmetry. We
conclude in section~\ref{sec:discussion}.  In addition, in
Appendix~\ref{app:spherical_equations} we show explicitly that our
equations system, in the spherical case, is manifestly regular.

%%%%%%%%%%%%%%%%%%%%%%%%%%%%%%%
%%%   EVOLUTION EQUATIONS   %%%
%%%%%%%%%%%%%%%%%%%%%%%%%%%%%%%

\section{Evolution equations}
\label{sec:evolution}

Since we are interested in finding a regularization algorithm that is
generic in the sense that it can be used with any formulation of the
evolution equations, we will introduce here two different systems of
evolution equations as test cases, namely the standard ADM system and
a strongly hyperbolic system.

%%%%%%%%%%%%%%%
%%%   ADM   %%%
%%%%%%%%%%%%%%%

\subsection{ADM evolution system}
\label{sec:ADM}

We will start from the standard ADM evolution equations in vacuum
\begin{eqnarray}
\frac{d}{dt} \: g_{ij} &=& -2 \alpha K_{ij} \; ,
\label{eq:gammadot} \\
\frac{d}{dt} \: K_{ij} &=& - \nabla_i \nabla_j \alpha \nonumber \\
&+& \alpha\, \Big( R_{ij} - 2 K_{il}\,{ K^l}_j + K\, K_{ij} \Big) \; ,
\label{eq:Kdot} 
\end{eqnarray}
where $\alpha$ is the lapse function, $\beta^i$ the shift vector,
$g_{ij}$ the spatial metric, $\nabla_i$ the covariant derivative
associated with $g_{ij}$, $K = g^{ij} K_{ij}$ the trace of the
extrinsic curvature and $R_{ij}$ the three-dimensional Ricci
tensor. In the above equations we have introduced the notation
$d/dt:=\partial/\partial t - {\cal L}_\beta$, with ${\cal L}_\beta$
the Lie derivative with respect to the shift.

These evolution equations are subject to the Hamiltonian and momentum
constraints, which in vacuum take the form
\begin{eqnarray}
H := R - K_{ij} K^{ij} + K^2 = 0 \; ,
\label{eq:ham} \\
M^i := \nabla_j \left( K^{ij} - g^{ij} K \right) =0 \; .
\label{eq:mom}
\end{eqnarray}
It is now well known that the ADM system of evolution equations
presented above is not strongly hyperbolic.  However, the question of
well-posedness is independent from the issue of the regularity of the
evolution system at the axis of symmetry, and in what follows we will
come back to the ADM system in order to show that the regularization
procedure proposed here will work for arbitrary formulations of the
evolution equations.  However, we will also introduce below a strongly
hyperbolic system.

%%%%%%%%%%%%%%%%%%%%%%%%%
%%%   HYPERBOLICITY   %%%
%%%%%%%%%%%%%%%%%%%%%%%%%

\subsection{Hyperbolic evolution system}
\label{sec:hyperbolic}

Having a well-posed system of evolution equations is crucial in order
to have a successful evolution code.  Many different well posed
formulations of the 3+1 evolution equations have been proposed in the
literature.  For simplicity, here we will follow the work of Nagy and
collaborators~\cite{Nagy:2004td}, but will adapt it to the case of
axial symmetry.

We start by defining the new dynamical quantities
\begin{equation}
\Delta^i := g^{mn} \Delta^i_{mn}
= g^{mn} \left( \Gamma^i_{mn} - \Gamma^i_{mn}\mid_{\rm flat}
\right) \,,
\label{eq:def_delta}
\end{equation}
with $\Gamma^i_{mn}$ the Christoffel symbols associated to the metric
$g_{ij}$ in some curvilinear coordinate system, and
$\Gamma^i_{mn}\mid_{\rm flat}$ the Christoffel symbols for flat space
in the same coordinate system.  As already mentioned
in~\cite{Alcubierre:2005gh}, the quantities $\Delta^i_{mn}$ are
components of a well defined tensor, while the $\Gamma^i_{mn}$ are not
and in fact are not even regular in spherical coordinates. One must
also remember that the contraction used to construct the vector
$\Delta^i=g^{mn} \Delta_{mn}$ must be done with the full metric
associated with the space under study, instead of the flat metric.

We will now promote the $\Delta^i$ to independent variables.  Using
the evolution equation~\eqref{eq:gammadot} we find the following
evolution equation for the vector $\Delta^i$
\begin{eqnarray}
\frac{d}{dt} \: \Delta^i &=& - \nabla_m \Big[ \alpha \left( 2 K^{im}
- g^{im} K \right) \Big] \nonumber \\
&+& 2\, \alpha\, K^{lm}\, \Delta^i_{lm} \; . 
\label{eq:Deltadot}
\end{eqnarray}

In order to study the hyperbolicity of the system, we must also say
something about the evolution of the gauge variables $\alpha$ and
$\beta^i$. For the lapse, we will choose a slicing condition of the
Bona--Masso family~\cite{Bona94b} of the form
\begin{equation}
\frac{d}{dt} \: \alpha = - \alpha^2 f(\alpha) \: K \; ,
\label{eq:BonaMasso}
\end{equation}
where $f(\alpha)$ is a positive but otherwise arbitrary function of
$\alpha$.  We will also assume that the shift vector is an {\em a
priory}\/ known function of spacetime $\beta^i(t,x^j)$, so that its
derivatives can be considered as source terms for the hyperbolicity
analysis.

Since we want to work with a first order system of equations, we
define the auxiliary variables
\begin{equation}
D_{ijk} := \frac{1}{2} \: \partial_i g_{jk} \,,
\qquad
F_i := \partial_i \ln \alpha
\, .
\label{eq:Def_DF}
\end{equation}

The evolution equations for $F_i$ and $D_{ijk}$ can be obtained
directly from~\eqref{eq:gammadot} and~\eqref{eq:BonaMasso}. Up to
principal part these evolution equations take the form
\begin{eqnarray}
\partial_0 D_{ijk} &\simeq& - \alpha \: \partial_i K_{jk} \; ,
\label{eq:Ddot} \\
\partial_0 F_i &\simeq& - \alpha f \: \partial_i K \; ,
\label{eq:Adot} 
\end{eqnarray}
where now $\partial_0 = \partial_t - \beta^i \partial_i$, and where
the symbol $\simeq$ indicates equal up to principal part.

In order to obtain a hyperbolic system, we will also modify the
evolution equations~\eqref{eq:Deltadot} for the vector $\Delta_i$ by
adding to them a multiple of the momentum constraints~\eqref{eq:mom}:
\begin{eqnarray}
\partial_0 \Delta_i &\simeq& - \alpha\,\Big( 2 \: \partial_m {K^m}_i -
\partial_i K \Big) + 2\, \alpha\, M_i \nonumber \\
&\simeq& - \alpha \: \partial_i K \; .
\end{eqnarray}

For the evolution equations for the extrinsic curvature $K_{ij}$ we
start by writing the Ricci tensor as
\begin{eqnarray}
R_{ij} &=& - \frac{1}{2}\, g^{lk}\, \partial_l \partial_kg_{ij}
+  g_{k(i} \partial_{j)}
\left\{ \Delta^k +
\Gamma^k\mid_{\rm flat}\right\}
\nonumber \\
&+&\,g^{l\,m}\,{\Gamma^k}_{i\,l} \Gamma_{k\,m\,j} + \frac{1}{2}
\,\partial_l g_{ij} \,
\left\{\Delta^l
+ \Gamma^l\mid_{\rm flat}\right\} \\
&+& \frac{1}{2}
\,g^{kl}g^{m n}
\left\{
\,2\, \partial_{(i}g_{ln} \partial_mg_{j)k}
-\partial_{(i}g_{ln}\partial_{j)}g_{km} \right\}\;, 
\nonumber
\label{eq:ricci}
\end{eqnarray}
where $\Gamma^l\mid_{\rm flat}\equiv g^{ij}\,\Gamma^l_{ij}\mid_{\rm
flat}$.  In the last expression the symmetrization refers to the $i,j$
indexes only.  The principal part of Ricci tensor becomes
\begin{eqnarray}
R_{ij} &\simeq& - \frac{1}{2}\, g^{lm} \partial_l \partial_m
g_{ij} + \partial_{(i} \Delta_{j)} \nonumber \\
&=& - \partial_m {D^m}_{ij} + \partial_{(i} \Delta_{j)} \; .
\hspace{8mm}
\label{eq:Ricciprincipal}
\end{eqnarray}
The evolution equation for $K_{ij}$ can then be written as
\begin{equation}
\partial_0 K_{ij} \simeq - \alpha \: \partial_k \Lambda^k_{ij} \; , 
\end{equation}
where we have defined
\begin{equation}
\Lambda^k_{ij} := {D^k}_{ij} + \delta^k_{(i} \left[ F_{j)} - \Delta_{j)}
\right]\,.
\label{eq:lambda2}
\end{equation}

Our system of evolution equations then takes the final form
\begin{eqnarray}
\partial_0 F_i &\simeq& - \alpha f \: \partial_i K \; ,
\label{eq:Adot2} \\
\partial_0 D_{ijk} &\simeq& - \alpha \: \partial_i K_{jk} \; ,
\label{eq:Ddot2} \\
\partial_0 \Delta_i &\simeq& - \alpha \: \partial_i K \; ,
\label{eq:Deltadot2} \\
\partial_0 K_{ij} &\simeq& - \alpha \: \partial_k \Lambda^k_{ij} \; .
\label{eq:Kdot2}
\end{eqnarray}

Even though the $\Lambda^k_{ij}$ are not independent quantities, it is
very useful for the subsequent analysis to write down their evolution
equations.  Using~\eqref{eq:Adot2}, \eqref{eq:Ddot2}
and~\eqref{eq:Deltadot2} we find
\begin{equation}
\partial_0 \Lambda^k_{ij} \simeq - \alpha \left[
g^{kl}\, \partial_l  K_{ij} + \left( f-1 \right) \delta^k_{(i}
\partial_{j)} K \right] \; .
\label{eq:lambdadot}
\end{equation}

We then have a system of 30 equations to study, corresponding to the 3
components of $F_i$, the 18 independent components of $D_{ijk}$, the 6
independent components of $K_{ij}$, and the 3 components of
$\Delta_i$. To proceed with the hyperbolicity analysis we will choose
a specific direction, say $x$, and ignore derivatives along the other
directions.  The idea is then to find 30 independent eigenfunctions
that will allow us to recover the 30 original quantities, where by
eigenfunctions here we mean linear combinations of the original
quantities $u=(F_i,D_{ijk},K_{ij},\Delta_i)$, of the form \mbox{$w_a =
\sum_b C_{ab} u_b$}, that up to principal part evolve as
\mbox{$\partial_t w_a + \lambda_a \partial_x w_a \simeq 0$}, with
$\lambda_a$ the corresponding eigenspeeds.

Taking then into account only derivatives along the $x$ direction we
immediately see that there are $14$ eigenfunctions that propagate
along the time lines with speed $-\beta^x$, namely $F_q$ and $D_{qij}$
for $q \neq x$.  Furthermore, taking $f$ times the trace
of~\eqref{eq:Ddot2} and subtracting it from~\eqref{eq:Adot2}, we find
that the 3 functions \mbox{$F_i - f {D_{im}}^m $} also propagate along
the time lines.  Finally, subtracting the trace of~\eqref{eq:Ddot2}
from~\eqref{eq:Deltadot2}, we find that the ${{D_i}^m}_m -\Delta_i$
are 3 more eigenfunctions that propagate along the time lines.  Thus,
we end up with $20$ eigenfunctions propagating along the time lines
with speed $-\beta^x$.

The remaining 10 eigenfunctions are obtained by combining the
evolution equation for the extrinsic curvature~\eqref{eq:Kdot2} with
the evolution equation for the ${\Lambda^q}_{ij}$, equation
~\eqref{eq:lambdadot}.  For simplicity, we assume that
$\beta^i=0$. Therefore, if $q\neq x$ we obtain the system
\begin{eqnarray}
\partial_0 K_{qi} &\simeq& - \alpha \: \partial_x \Lambda^x_{qi} \; , \\
\partial_0 \Lambda^x_{qi} &\simeq& - \alpha\, g^{xx} \partial_x K_{qi} \; ,
\end{eqnarray}
from which is clear that we have 8 new eigenfunctions of the form
\begin{equation}
\sqrt{g^{xx}}\, K_{qi} \mp \Lambda^x_{qi} \; ,
\end{equation}
with characteristic speed given by $\pm \alpha\, \sqrt{g^{xx}}$.
Finally, taking the trace of the extrinsic curvature and of
$\Lambda^k_{ij}$, we find
\begin{eqnarray}
\partial_0 K &\simeq& - \alpha \: \partial_x \Lambda^x \; , \\
\partial_0 \Lambda^x &\simeq& - \alpha\, f\, g^{xx} \partial_x K \; ,
\end{eqnarray}
with $\Lambda^x := g^{mn} \Lambda^x_{mn}$. So that our final pair
of eigenfunctions are
\begin{equation}
\sqrt{f\, g^{xx}}\, K \mp \Lambda^x \; ,
\end{equation}
with characteristic speed $\pm \alpha\, \sqrt{f\, g^{xx}}$.

In this way we see that for the evolution system where the vector
$\Delta^i$ has been promoted to an independent variable, and a
multiple of the momentum constraint has been added to its evolution
equation, one can obtain a complete set of independent eigenfunctions,
showing that the system is indeed strongly hyperbolic.

%%%%%%%%%%%%%%%%%%%%%%%%%%%%%%
%%%   SPHERICAL SYMMETRY   %%%
%%%%%%%%%%%%%%%%%%%%%%%%%%%%%%

\section{Regularity in spherical symmetry}
\label{sec:spherical}

%%%%%%%%%%%%%%%%%%%%%%%%%%%%%%%%%%%%%%%%
%%%   PARITY IN SPHERICAL SYMMETRY   %%%
%%%%%%%%%%%%%%%%%%%%%%%%%%%%%%%%%%%%%%%%

\subsection{Parity conditions}
\label{sec:parityspherical}

There are in fact two types of regularity conditions for the metric
components.  One set of conditions comes directly from symmetry
considerations.  In spherical symmetry we can write the metric quite
generally as
\begin{eqnarray}
ds^2 &=& - \left( \alpha - \beta_r \beta^r \right) dt^2 + 2\, \beta_r dr
dt \nonumber \\ && +\, g_{rr} dr^2 + g_{\theta \theta}\, d \Omega^2 \; ,
\label{eq:metricspherical}
\end{eqnarray}
where $\alpha$, $\beta^r$, $g_{rr}$ and $g_{\theta \theta}$ are
functions of $r$ and $t$ only, and $d \Omega^2$ is the solid angle
element: $d\Omega^2= d\theta+\sin\theta\,d\phi^2$.  Spherical symmetry
means that a reflection through the origin should leave the metric
unchanged.  By making the transformation \mbox{$r \rightarrow -r$} in
the above metric we see that this implies that
\begin{eqnarray}
\alpha(r) &=& \alpha(-r) \; , \\
\beta^r(r) &=& - \beta^r(-r) \; , \\
g_{rr}(r) &=& g_{rr}(-r) \; , \\
g_{\theta \theta}(r) &=& g_{\theta \theta}(-r) \; ,
\end{eqnarray}
or in other words, $\alpha$, $g_{rr}$ and $g_{\theta \theta}$ must be
even functions of $r$, while $\beta^r$ must be odd.  The parity of the
spatial metric coefficients clearly must be inherited by the
corresponding components of the extrinsic curvature, so that $K_{rr}$
and $K_{\theta \theta}$ must also be even functions of $r$.

%%%%%%%%%%%%%%%%%%%%%%%%%%%%%%%%%%%%%%%%%%%%%%%%
%%%   LOCAL FLATNESS IN SPHERICAL SYMMETRY   %%%
%%%%%%%%%%%%%%%%%%%%%%%%%%%%%%%%%%%%%%%%%%%%%%%%

\subsection{Local flatness}

Parity considerations are not enough in order to have a regular
evolution.  There are extra regularity conditions that the geometric
variables $(g_{ij},K_{ij})$ have to satisfy at the origin that are a
consequence of the fact that the manifold must be locally flat.

Local flatness implies that close to the origin one should be able to
write the spatial metric as
\begin{equation}
dl^2 = d\tilde{r}^2 + \tilde{r}^2 d\Omega^2 \; ,
\end{equation}
with $\tilde{r}$ a radial coordinate that measures proper distance
from the origin. If we now change the radial coordinate to some new
coordinate $r$ related to $\tilde{r}$ through
$\tilde{r}=\tilde{r}(r)$, the metric will transform into
\begin{equation}
dl^2 = \left( \frac{d\tilde{r}}{dr} \right)^2 dr^2
+ r^2 \left( \frac{\tilde{r}}{r} \right)^2 d\Omega^2 \; .
\end{equation}
Expanding now $\tilde{r}$ in Taylor around the origin we find
\begin{equation}
\tilde{r} \simeq r \left( \frac{d\tilde{r}}{dr} \right)_{r=0} \; ,
\end{equation}
so that close to the origin we will have
\begin{equation}
dl^2 = \left( \frac{d\tilde{r}}{dr} \right)^2_{r=0}
\left( dr^2 + r^2 d\Omega^2 \right) \; .
\end{equation}
In other words, for any arbitrary radial coordinate $r$ the metric at
the origin must be proportional to the flat metric ({\em i.e.} it must
be conformally flat). Taking this result together with the parity
conditions derived in the last section we see that we can rewrite the
spatial metric in spherical symmetry as
\begin{equation}
dl^2 = A dr^2 + r^2 T d\Omega^2 \; ,
\label{eq:metricspherical1}
\end{equation}
where $A$ and $T$ are such that close to the origin
\begin{equation}
A = A_0 + r^2 A_1 \; , \qquad T = T_0 + r^2 T_1 \; , 
\end{equation}
with $A_0=T_0$ functions of $t$ only.

The results just described where in fact already presented
in~\cite{Alcubierre04a}.  In that reference the condition that
\mbox{$A_0(t)=T_0(t)$} was implemented by defining a new dynamical
variable that is odd at the origin:
\begin{equation}
\lambda := \frac{1}{r} \left( 1 - \frac{A}{T} \right) \; ,
\label{eq:lambda}
\end{equation}
and deriving an evolution equation for it.  Such a regularization
procedure works well in spherical symmetry, but its direct
generalization to the case of axial symmetry has one very serious
drawback.  The problem arises because such an algorithm introduces
terms of the form $\partial_z \lambda/\rho$, with $\rho$ and $z$
cylindrical coordinates, that change the characteristic structure of
the evolution equations and can therefore spoil the hyperbolicity of a
given formulation.

Because of this, we will introduce here a different regularization
procedure that can be generalized more directly to the case of
axi-symmetry.  Let us start by defining the variables
\begin{equation}
H := \frac{A+T}{2} \; , \qquad J := \frac{A-T}{2r^2} \; .
\end{equation}
The results derived above imply that both $H$ and $J$ are regular
functions that are even at the origin.  The definitions of $H$ and $J$
can easily be inverted to give
\begin{equation}
A := H + r^2 J \; , \qquad T := H - r^2 J \; ,
\end{equation}
so that the spatial metric can be rewritten as
\begin{equation}
dl^2 = \left( H + r^2 J \right) dr^2
 + r^2 \left( H - r^2 J \right) d\Omega^2 \; .
\label{eq:metric_esf_HJ}
\end{equation}

In order for this form of the metric to be maintained in time, one
must ask for the extrinsic curvature to behave in the same way.  We
will then take the extrinsic curvature to be
\begin{equation}
K_{ij} = \left( \begin{array}{c c c}
K_A & 0       & 0 \\
0   & r^2 K_T & 0 \\
0   & 0       & r^2 K_T \sin^2 \theta
\end{array} \right) \; ,
\end{equation}
where $K_A \equiv K_H + r^2 K_J$ and $K_T \equiv K_H - r^2 K_J$, and
with $(K_H, K_J)$ even functions at the origin.

The $\Delta^i$ vector in this case takes the simple form
\mbox{$\Delta^i=\left(\Delta^r(t,r),0,0\right)$}, where
\begin{eqnarray}
\Delta^r = \frac{1}{A} \Bigg(\frac{D_{rrr}}{A}
- \frac{2\,\left(D_{r\theta\theta} -2\,r\,J\right)}{T}
\Bigg) \; .
\label{eq:deltar}
\end{eqnarray}
In this last expression we used the definition~\eqref{eq:Def_DF} for
the spatial derivatives. The parity properties of $\Delta^r$ follow
directly from those of the metric, and one finds that $\Delta^r$ must
be odd at the origin.

%%%%%%%%%%%%%%%%%%%%%%%%%%%%%%%%%%%%%%%%%%%%%%%%
%%%   REGULARIZATION IN SPHERICAL SYMMETRY   %%%
%%%%%%%%%%%%%%%%%%%%%%%%%%%%%%%%%%%%%%%%%%%%%%%%

\subsection{Regularization algorithm}

The main idea of the regularization algorithm is simply to evolve
directly the variables $(H,J,\partial_r H,\partial_r
J,K_H,K_J,\Delta^r)$ imposing the appropriate parity conditions on
these variables, which will automatically guarantee that local
flatness is maintained.

The parity conditions are in fact very easy to implement numerically.
The easiest way to do this is to stagger the origin, with a fictitious
grid point located at $r= -\Delta r/2$.  One then implements the
parity conditions across the origin by simply copying the value of a
given variable from $r=\Delta r/2$ to $r=-\Delta r/2$, with the
appropriate sign.

The evolution equations for $H$, $J$, $\partial_rH$ and $\partial_rJ$
are in fact trivial to obtain. For example, in the case of zero shift,
they have the form
\begin{eqnarray}
\partial_t H &=& - 2 \,\alpha\, K_H \; , \\
\partial_t J &=& - 2 \,\alpha\, K_J \; , \\
\partial_t D_H &=& - 2\, \partial_r\Big(\alpha\, K_H\Big) \; , \\
\partial_t D_J &=& - 2\, \partial_r\Big(\alpha\, K_J\Big) \; ,
\label{eq:S_variables}
\end{eqnarray}
where we defined $\partial_rH\equiv D_H$ and $\partial_rJ\equiv D_J$.
The evolution equations for $K_H$ and $K_J$ can also be obtained
directly from those of $K_A$ and $K_T$.  The resulting equations are
again trivial to derive but rather long, and we will write them
explicitly in the Appendix~\ref{app:spherical_equations}.  However,
the evolution equation for $K_H$ looks like
\begin{eqnarray}
\partial_t K_H &=& 
\frac{\alpha\,H^3}{2\,r\,A^2\,T^2}\,
\Big(\Delta^r\,H^2-F_r\,H-2\,D_H\Big)
\nonumber \\
&+& {\cal H}\,,
\label{eq:KH_dot}
\end{eqnarray}
where ${\cal H}$ stands for terms that are not divided by $r$.  By
simple inspection one can see that all terms in the evolution equation
for $K_H$ are manifestly regular. The evolution equation for $K_J$, on
the other hand, takes the form
\begin{eqnarray}
\partial_t K_J &=& -\frac{\alpha\,H^4\,
\Big(H\,\Delta^r - F_r\Big)}{2\, r^3\, A^2\,T^2}
+ \frac{\alpha\,H^4\,\Big(H\,\partial_r\Delta^r
-\partial_rF_r\Big)}{2\, r^2\,A^2\, T^2} \nonumber \\
&+& {\cal J}\; ,
\label{eq:KJdot}
\end{eqnarray}
where $\cal{J}$ stands for terms that either have no divisions by $r$,
or else involve terms of the form $(D_H)^2/r^2$, $D_J/r$, etc., which
are manifestly regular.

In the above equation, one must remember that because of the behavior
of $\Delta^r$ and $F_r$, $\partial_r\Delta^r$ and $\partial_rF_r$ are
even functions at the origin. One can now see that the first two terms
in~\eqref{eq:KJdot} are regular by first noticing that they can be
joined in pairs to form a single derivative, so that the equation
becomes
\begin{eqnarray}
\partial_t K_J &=& \frac{\alpha\, H^5}{2\,r\,A^2\,T^2} \: \partial_r
\left( \frac{\Delta^r}{r} \right) - \frac{\alpha\,H^4}{2\, r\, A^2\, T^2} \:
\partial_r \left( \frac{F_r}{r} \right) \nonumber \\
&+& {\cal J}\; .
\label{eq:KJdot_regular}
\end{eqnarray}
It is now easy to see that this last evolution equation is manifestly
regular, due to the fact that \mbox{$\Delta^r/r \sim {\rm constant} +
O(r^2)$}, so that $\partial_r \left({\Delta^r}/{r} \right)\sim O(r)$,
and $F_r/r \sim {\rm constant}+O(r^2)$, so that $\partial_r
(F_r/r)\sim O(r)$.

One can also see that the evolution equation for $\Delta^r$, and both
the Hamiltonian and momentum constraints, are trivially regular. On
the other hand, if one uses the regularization procedure
of~\cite{Alcubierre04a}, the momentum constraint remains irregular.

%%%%%%%%%%%%%%%%%%%%%%%%%%
%%%   AXIAL SYMMETRY   %%%
%%%%%%%%%%%%%%%%%%%%%%%%%%

\section{regularity in axial symmetry}
\label{sec:axial}

%%%%%%%%%%%%%%%%%%%%%%%%%%%%%%%%%%%%
%%%   PARITY IN AXIAL SYMMETRY   %%%
%%%%%%%%%%%%%%%%%%%%%%%%%%%%%%%%%%%%

\subsection{Parity conditions}
\label{sec:parityaxial}

In the case of axial symmetry, the spacetime metric can be written in
cylindrical coordinates $(\rho,z,\phi)$ as
\begin{eqnarray}
ds^2 &=& - \left( \alpha - \beta_i \beta^i \right) dt^2
+ 2 \left( \beta_\rho d\rho + \beta_z dz + \beta_\phi d\phi \right) dt
\nonumber \\
&+& g_{\rho \rho} d\rho^2 + g_{zz} dz^2 + g_{\phi \phi} d\phi^2 \nonumber \\
&+& 2 \left(  g_{\rho z} d\rho dz + g_{\rho \phi} d\rho d\phi +
g_{z \phi} dz d\phi \right) \; .
\label{eq:metricaxial}
\end{eqnarray}
As before, axial symmetry implies that the metric should remain
unchanged under the transformation \mbox{$\rho \rightarrow -\rho$},
which implies
\begin{eqnarray}
\alpha(\rho) &=& \alpha(-\rho) \; , \\
\beta_\rho(\rho) &=& - \beta_\rho(-\rho) \; , \\
\beta_z(\rho) &=& \beta_z(-\rho) \; , \\
\beta_\phi(\rho) &=& \beta_\phi(-\rho) \; , \\
g_{\rho \rho}(\rho) &=& g_{\rho \rho}(-\rho) \; , \\
g_{z z}(\rho) &=& g_{z z}(-\rho) \; , \\
g_{\phi \phi}(\rho) &=& g_{\phi \phi}(-\rho) \; , \\
g_{\rho z}(\rho) &=& - g_{\rho z}(-\rho) \; , \\
g_{\rho \phi}(\rho) &=& - g_{\rho \phi}(-\rho) \; , \\
g_{z \phi}(\rho) &=& g_{z \phi}(-\rho) \; .
\end{eqnarray}
Again, the components of the extrinsic curvature inherit their parity
properties from the corresponding metric coefficients.

%%%%%%%%%%%%%%%%%%%%%%%%%%%%%%%%%%%%%%%%%%%%
%%%   LOCAL FLATNESS IN AXIAL SYMMETRY   %%%
%%%%%%%%%%%%%%%%%%%%%%%%%%%%%%%%%%%%%%%%%%%%

\subsection{Local flatness}

As in the spherical case, parity conditions are not enough. One also
needs to consider the conditions arising from the fact that space must
be locally flat at the axis of symmetry.  We will derive those
conditions here somewhat informally in order to have a more intuitive
idea of where they come from.  For a more formal proof the reader can
look at~\cite{Rinne:2005sk}, where the same conditions are arrived at
by solving the Killing equation for axial symmetry.

Let us start by considering the general spatial metric in Cartesian
coordinates
\begin{eqnarray}
dl^2 &=& g_{xx} dx^2 + g_{yy} dy^2 + g_{zz} dz^2 \nonumber \\
&+& 2 g_{xy} dx dy + 2 g_{xz} dx dz + 2 g_{yz} dy dz \; .
\end{eqnarray}
Axial symmetry implies, in particular, that the metric must be
invariant under reflections about the $x$ and $y$ axes, and under
exchange of $x$ for $y$.  Local flatness also implies that the metric
must be smooth.  These two requirements together imply that for fixed
$z$ we must have
\begin{eqnarray}
g_{xx} &\sim& k_\rho + {\cal O}(x^2+y^2)
\sim k_\rho + {\cal O}(\rho^2) \; , \\
g_{yy} &\sim& k_\rho + {\cal O}(x^2+y^2)
\sim k_\rho + {\cal O}(\rho^2) \; , \\
g_{zz} &\sim& k_z + {\cal O}(x^2+y^2)
\sim k_z + {\cal O}(\rho^2) \; , \\
g_{xy} &\sim& {\cal O}(xy) \sim {\cal O}(\rho^2) \; , \\
g_{xz} &\sim& {\cal O}(x) \sim {\cal O}(\rho) \; , \\
g_{yz} &\sim& {\cal O}(y) \sim {\cal O}(\rho) \; ,
\end{eqnarray}
where $k_\rho$ and $k_z$ are constants. Let us now consider a
transformation to cylindrical coordinates $(\rho,z,\phi)$:
\begin{equation}
x = \rho \cos \phi \; , \qquad y = \rho \sin \phi \; , \qquad z=z \; .
\end{equation}
Under such a transformation we have
\begin{eqnarray}
g_{\rho \rho} &=& g_{xx} \cos^2 \phi + g_{yy} \sin^2 \phi \nonumber \\
&& + 2 g_{xy} \sin \phi \cos \phi \; , \\
g_{zz} &=& g_{zz} \; , \\
g_{\phi \phi} &=& \rho^2 \left( g_{xx} \sin^2 \phi + g_{yy} \cos^2 \phi
\right. \nonumber \\
&& \left. - 2 g_{xy} \sin \phi \cos \phi \right) \; , \\
g_{\rho z} &=& g_{xz} \cos \phi + g_{yz} \sin \phi \; , \\
g_{\rho \phi} &=& \rho \left( g_{yy} - g_{xx} \right) \sin \phi \cos \phi
\nonumber \\
&& + \rho \: g_{xy} \left( \cos^2 \phi - \sin^2 \phi \right) \; , \\
g_{z \phi} &=& \rho \left( - g_{xz} \sin \phi + g_{yz} \cos \phi \right) \; .
\end{eqnarray}

From the behavior of the different Cartesian metric components near
the axis we then see that
\begin{eqnarray}
g_{\rho \rho} &\sim& k_\rho + {\cal O}(\rho^2) \; , \\
g{zz} &\sim& k_z + {\cal O}(\rho^2) \; , \\
g_{\phi \phi} &\sim& \rho^2 \left( k_\rho + {\cal O}(\rho^2) \right) \; , \\
g_{\rho z} &\sim& {\cal O}(\rho) \; , \\
g_{\rho \phi} &\sim& {\cal O}(\rho^3) \; , \\
g_{z \phi} &\sim& {\cal O}(\rho^2) \; .
\end{eqnarray}
Therefore the spatial metric can be written as
\begin{eqnarray}
dl^2 &=& A\,d\rho^2 + B\,dz^2 +\rho^2 T d\phi^2 
+ 2\,\Big(\rho\, C\, d\rho\,dz
\nonumber \\
&+&
\rho^3\, C_1\, d\rho\, d\phi
\,+\, \rho^2\, C_2\, dz\, d\phi\Big)\,,
\label{eq:metric_axig2}
\end{eqnarray}
with $(A,B,T,C,C_1,C_2)$ all even functions of $\rho$ on the axis.
Again, let us define the new variables
\begin{equation}
H := \frac{A+T}{2} \; , \qquad J := \frac{A-T}{2\rho^2} \; .
\end{equation}
The results derived above imply that both $H$ and $J$ are regular
functions that are even in $\rho$.  The definitions of $H$ and $J$ can
easily be inverted to give
\begin{equation}
A := H + \rho^2 J \; , \qquad T := H - \rho^2 J \; ,
\end{equation}
so that the spatial metric~\eqref{eq:metric_axig2} can be rewritten as
\begin{eqnarray}
dl^2 &=& (H +\rho^2J)\,d\rho^2 + B\,dz^2 +\rho^2 (H-\rho^2J)\, d\phi^2
\nonumber \\
&+& 2\, \left( \rho\, C\, d\rho\, dz\, +\,\rho^3\, C_1 
\, d\rho\, d\phi
\,+\, \rho^2\, C_2\, dz\, d\phi \right) . \hspace{10mm} 
\label{eq:metric_axig}
\end{eqnarray}

For the extrinsic curvature $K_{ij}$ we take the similar form:
\begin{equation}
K_{ij} =
\left(
\begin{array}{c c c}
K_A, & \rho\,K_C & \rho^3\,K_{C_1} \\
\rho\,K_C & K_B & \rho^2\,K_{C_2} \\
\rho^3\,K_{C_1} & \rho^2\,K_{C_2} &\rho^2\,K_T\,
\end{array} \right) \; ,
\end{equation}
with $K_A\,=\,K_H\,+\,\rho^2\,K_J\,, K_T\,=\,K_H\,-\,\rho^2\,K_J$.
The extrinsic curvature components are given in such a way that all
the functions are even, as in the metric case.

The $\Delta^i$ vector takes the form
$(\Delta^\rho,\Delta^z,\Delta^\phi)$, and is a well defined
vector. The general expression for $\Delta^i$ can be obtained directly
from its definition. In this way, we find that $\Delta^\rho$ is odd,
while $\Delta^z$ and $\Delta^\phi$ are even with respect to
reflections on the axis.

%%%%%%%%%%%%%%%%%%%%%%%%%%%%%%%%%%%%%%%%
\subsection{Regularization algorithm}
%%%%%%%%%%%%%%%%%%%%%%%%%%%%%%%%%%%%%%%%

The main idea of the regularization algorithm is again to evolve
directly $(H,J,D_\rho H,D_\rho J,D_zH,D_zJ,K_H,K_J)$, instead of
$(A,T,D_{\rho\rho\rho},D_{\rho zz},D_{z\rho\rho },D_{zzz},K_A,K_T)$,
together with the other metric and extrinsic curvature coefficients
and the $\Delta^i$.  The corresponding parity conditions can again be
implemented numerically by staggering the axis with a fictitious grid
point located at $\rho= -\Delta\rho/2$.

The evolution equations for $K_H$ and $K_J$ can again be obtained
directly from those of $K_A$ and $K_T$.  The resulting equations are
very long so we will not write them here, but they are again trivial
to obtain. Consider, for example, the case of the hyperbolic system
without rotation and, by simplicity, shift vanish. That is, equations
\eqref{eq:gammadot}, \eqref{eq:Kdot}, \eqref{eq:Deltadot} and
\eqref{eq:BonaMasso} with $C_1=C_2=0$. In this case the evolution
equation for $K_H$ is manifestly regular, but the evolution equation
for $K_J$ has terms that at first sight appear irregular and have the
form
\begin{eqnarray}
\partial_tK_J &=& - \frac{\alpha B^2 H^4}{2  \rho T^2 (A B-\rho^2 C^2)^2}
\Bigg(\frac{H \Delta^\rho}{\rho^2}
- \frac{H \partial_\rho \Delta^\rho}{\rho}  \nonumber \\
 &-& \frac{F_\rho}{\rho^2} + \frac{\partial_\rho F_\rho}{\rho}
\Bigg) + {\cal{J}} \; ,
\label{eq:KJ_rho}
\end{eqnarray}
where again $\cal{J}$ stands for terms that either involve no
divisions by $\rho$, or involve terms like $(D_\rho H\,\partial_\rho
F_\rho)/\rho^2$, $D_\rho J/\rho$, which are manifestly regular.

Just as in the spherical case, one can see that the first terms
in~\eqref{eq:KJ_rho} are regular by noticing that they can be joined
in pairs to form a single derivative, so that the evolution equation
for $K_J$ becomes
\begin{eqnarray}
\partial_tK_J &=& \frac{\alpha\, B^2\, H^4}
{2\,\rho\,T^2\,(A B - \rho^2\, C^2)^2}
\Bigg(H\, \partial_\rho\left(\frac{\Delta^{\rho}}{\rho}\right)
\nonumber \\
&-& \partial_\rho\left(\frac{F_\rho}{\rho}\right)\Bigg) + \cal{J} \; .
\label{eq:KJ_rho_regular}
\end{eqnarray}
It is now easy to see that this last evolution equation is regular,
due to the fact that $\Delta^\rho/\rho \sim {\rm constant} +
O(\rho^2)$, so that, $\partial_\rho \left(\Delta^\rho/{\rho}
\right)\sim O(\rho)$, and $F_\rho/\rho \sim {\rm constant}+O(\rho^2)$,
so that $\partial_\rho (F_\rho/\rho)\sim O(\rho)$. On the other hand,
by inspection one can see that all terms in the remaining evolution
equations are manifestly regular leaving us with a regular system of
equations for the axial symmetric case. As final comment, notice that
since the regularization algorithm is very general, one can use it in
order to have a regularized system in the ADM case. One obtain the
similar evolution equation as in the hyperbolic system.  For example,
the evolution equation without rotation and shift vanish for $K_J$
looks like,
\begin{eqnarray}
\partial_tK_J &=& \frac{\alpha\, B\, H^4}
{2\,\rho\,A^2\,T^2\,(A B - \rho^2\, C^2)^2}
\Bigg(\partial_\rho\left(\frac{D_{\rho zz}}{\rho}\right)
\nonumber \\
&-&B\, \partial_\rho\left(\frac{F_{\rho}}{\rho}\right)\Bigg) + \cal{J} \;.
\label{eq:ADMKJ_rho_regular}
\end{eqnarray}
One can see that the last equation is regular on the axis.

%%%%%%%%%%%%%%%%%%%%
%%%   EXAMPLES   %%%
%%%%%%%%%%%%%%%%%%%%

\section{Examples}
\label{sec:examples}

In the simulations shown below we will see how the regularization
procedure described in the previous sections works in practice.  We
will consider first an evolution of Minkowski spacetime with a
non-trivial slicing in order to compare with the algorithm presented
in~\cite{Alcubierre04a}. We will perform similar simulations using
both a spherically symmetric and an axially symmetric code. Also, in
order to see that the regularization procedure is independent of the
hyperbolicity of the system of evolution equations, we will do the
axi-symmetric simulation using both the ADM system and the strongly
hyperbolic system derived in Section~\ref{sec:hyperbolic}.  As a
second example, we will consider a Brill wave spacetime as a
non-trivial test of the regularization procedure in axi-symmetry.

All runs have been performed using a method of lines with iterative
Crank-Nicholson integration in the time, and standard second-order
centered differences in space.

%%%%%%%%%%%%%%%%%%%%%%%%%%%%%
%%%   SPHERICAL EXAMPLE   %%%
%%%%%%%%%%%%%%%%%%%%%%%%%%%%%

\subsection{Minkowski in spherical symmetry}

As a first example of the regularization method we evolve Minkowski
spacetime with a non-trivial slicing and vanishing shift, using the
hyperbolic system presented in Section~\ref{sec:hyperbolic}. The
initial data corresponds to a trivial slice so that
\begin{eqnarray}
\hspace{-8mm} A &=& T = 1 \; , \\
K_A &=& K_T = 0 \; ,
\end{eqnarray}
which implies,
\begin{eqnarray}
H &=& 1 \;, \quad J=0 \; .\\
K_H &=& K_J =0 \; .
\end{eqnarray}
In order to have a non-trivial evolution, we chose a non-trivial
initial lapse profile of the form:
\begin{eqnarray}
\alpha(t=0) &=& 1 + r^2 R \left( \exp \left[ -\left( \frac{r - r_0}{\sigma}
\right)^2 \right]\right.
\nonumber\\ 
&+& \left. \exp \left[ -\left( \frac{r + r_0}{\sigma} \right)^2 \right]
\right) \; .
\label{eq:lapsespher} 
\end{eqnarray}
This specific lapse profile has been chosen because it guarantees that
$\alpha(t=0)$ is regular at the origin. In the simulation shown below
we have taken the Gaussian parameters to be $R=0.001$, $r_0=5.0$ and
$\sigma=1.0$. We will evolve the lapse using a Bona-Masso slicing
condition but restricted to harmonic slicing, that is $f(\alpha)=1$.
Furthermore, we have used a grid spacing of $\Delta r=0.1$ with the
outer boundaries at $r=100$, and a Courant factor of $\Delta t /
\Delta r = 0.5$.

\begin{figure}
\epsfig{file=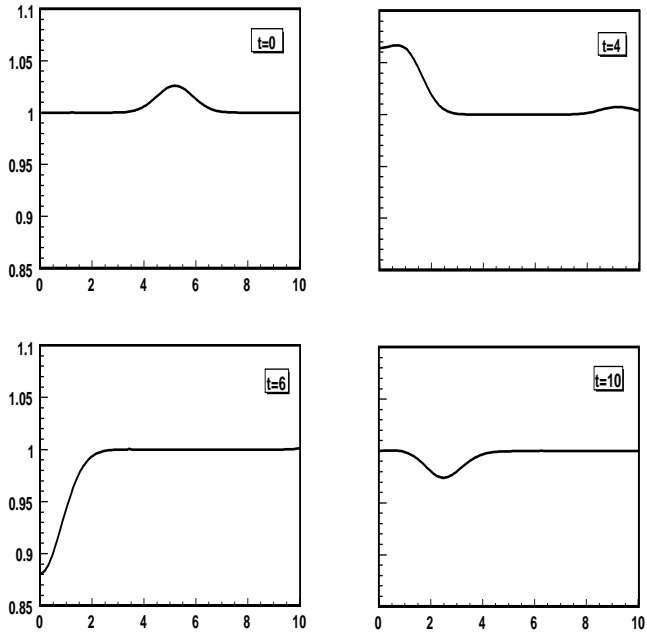,width=90mm,height=90mm}
\caption{Evolution of Minkowski spacetime with a non-trivial slicing
using a spherically symmetric code. The plots show the evolution of
the lapse function $\alpha$ at different times.  Notice how smoothly
the lapse behaves as the Gaussian pulse goes through the origin.}
\label{fig:alpha_sphe} 
\end{figure}

Figure~\ref{fig:alpha_sphe} shows the evolution of the lapse function
$\alpha$ near the origin. One can clearly see that the lapse remains
perfectly smooth when the Gaussian pulse goes through the origin.  The
system can in fact evolve for very long times and the behavior at the
origin remains well behaved.

\subsection{Minkowski in axial symmetry using ADM}

In this section we present a similar evolution to the one of the last
Section, but done now with an axi-symmetric code using the ADM
formulation.  We again consider initial data corresponding to a
trivial slice of Minkowski, so that the initial metric and extrinsic
curvature have the form
\begin{eqnarray}
A &=& B = T= 1 \; ,\\
C &=& C_1 = C_2 = 0 \; ,\\
K_A &=& K_B = K_T =0 \; ,\\
K_C &=& K_{C_1} = K_{C_2}=0 \; ,\\
\end{eqnarray}
which implies,
\begin{eqnarray}
H &=& 1 \; , \quad J=0 \; , \\
K_H &=& K_J = 0 \; ,
\end{eqnarray}
Again, we chose an initial non-trivial lapse profile which for
simplicity is now centered at the origin:
\begin{equation}
\alpha(t=0) = 1 + R \exp\left[-\left(\frac{z}{\sigma_z}\right)^2 
- \left(\frac{\rho}{\sigma_\rho}\right)^2 \right] \; .
\label{eq:lapseaxial} 
\end{equation}

For the particular simulation presented here we have taken $R=0.015$,
$\sigma_\rho=\sigma_z=2.5$. We will again evolve the lapse using
harmonic slicing, $f(\alpha)=1$. For this simulation we have used a
grid spacing of $\Delta \rho=\Delta z=0.125$, with a Courant factor of
$\Delta t / \Delta \rho = 0.25$.  The outer boundaries are at
$\rho=129,z=\pm 129$. Furthermore, Kreiss-Oliger fourth order
dissipation has been added for stability~\cite{Gustafsson95}, whereby
we modify a given variable $u$ in the new time-step by adding to its
evolution equation
\begin{eqnarray}
\partial_t u_{i,j} &\rightarrow& \partial_t u_{i,j}
- \frac{\epsilon}{\Delta \rho} \left( u_{i+2,j} - 4 u_{i-1,j} \right.
\nonumber \\
&& \left. + 6 u_{i,j} - 4 u_{i-1,j} + u_{i-2,j} \right) \nonumber \\
&& - \frac{\epsilon}{\Delta z} \left( u_{i,j+2} - 4 u_{i,j+2} \right.
\nonumber \\
&& \left. + 6 u_{i,j} - 4 u_{i,j-1} + u_{i,j-2} \right) ,
\label{eq:diss}
\end{eqnarray}
where the indices $i,j$ refer to the grid points along the $\rho$ and
$z$ directions.  For this simulation the dissipation coefficients have
been taken to be $\epsilon_\rho=\epsilon_z=0.04$.

\begin{figure}
\epsfig{file=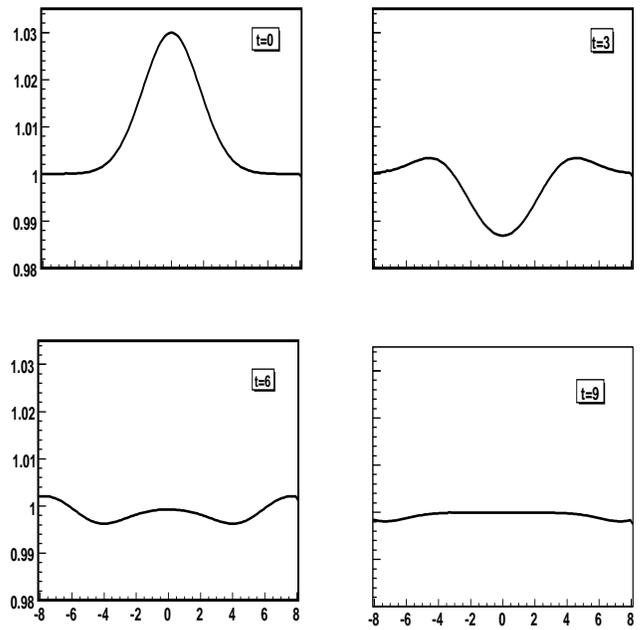,width=90mm,height=90mm}
\caption{Evolution of Minkowski spacetime with a non-trivial slicing
using an axi-symmetric code with the ADM formulation. The plots show
the evolution of the lapse function $\alpha$ at different times along
the $\rho$ axis.}
\label{fig:ADM_lapse} 
\end{figure}

Figure~\ref{fig:ADM_lapse} shows the evolution of lapse function
$\alpha$ along the $\rho$ axis. Notice that again there is no problem
at the axis of symmetry: The lapse evolves as a wave, goes through the
origin, and finally returns to 1.  The evolution time is only limited
by the instabilities produced from the fact that ADM is not strongly
hyperbolic.

\subsection{Minkowski in axial symmetry using a hyperbolic formulation}

In our next example, we consider exactly the same situation as in the
last Section but using now a hyperbolic formulation. As before, we
have used a grid spacing of $\Delta\rho=\Delta z=0.125$ and a Courant
factor of $\Delta t / \Delta \rho=0.25$. Again, Kreiss-Oliger second
order dissipation has been added for stability with dissipation
coefficients $\epsilon_\rho=\epsilon_z=0.04$.

\begin{figure}
\epsfig{file=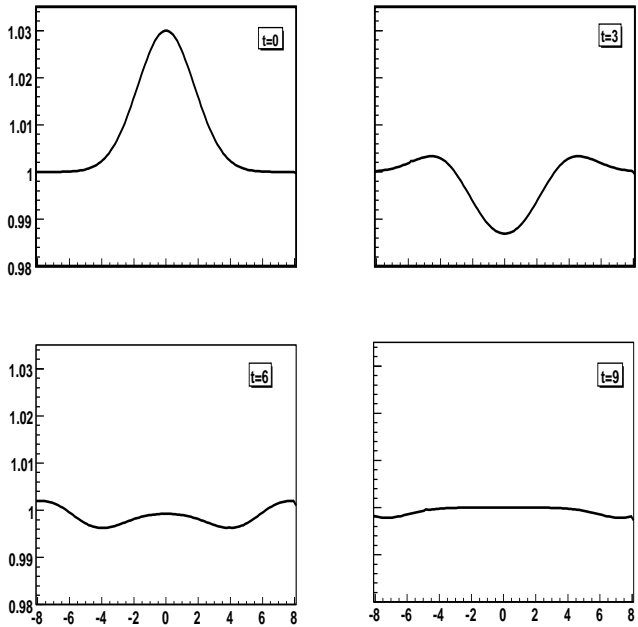,width=90mm,height=90mm}
\caption{Evolution of Minkowski spacetime with a non-trivial slicing
using an axi-symmetric code with a hyperbolic formulation. The plots
show the evolution of the lapse function $\alpha$ at different times
along the $\rho$ axis.}
\label{fig:hy_alpha} 
\end{figure}

\begin{figure}
\epsfig{file=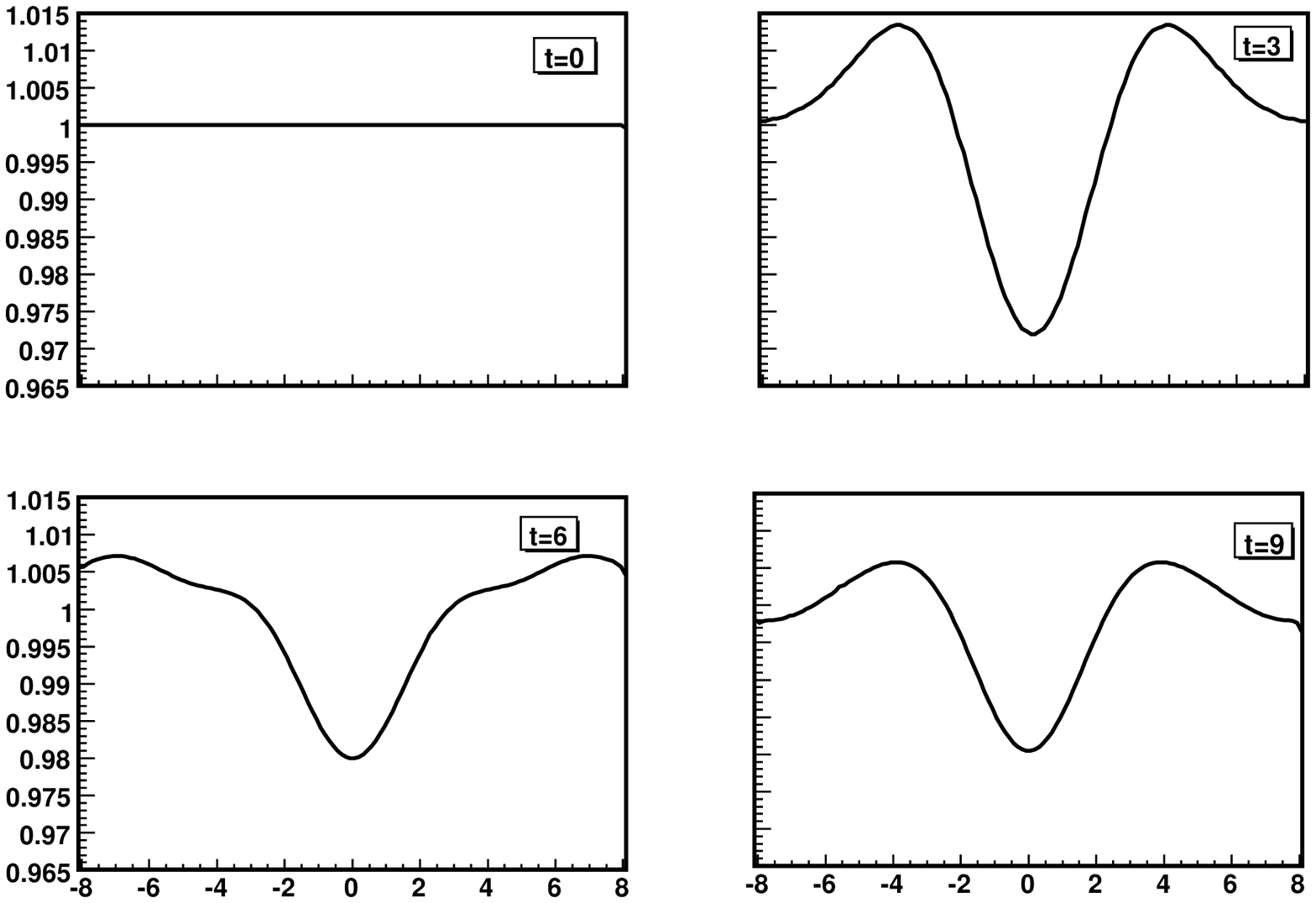,width=90mm,height=90mm}
\caption{Evolution of Minkowski spacetime with a non-trivial slicing
using an axi-symmetric code with a hyperbolic formulation. The plots
show the evolution of the metric function $A$ at different times
along the $\rho$ axis.}
\label{fig:hy_KA} 
\end{figure}

\begin{figure}
\epsfig{file=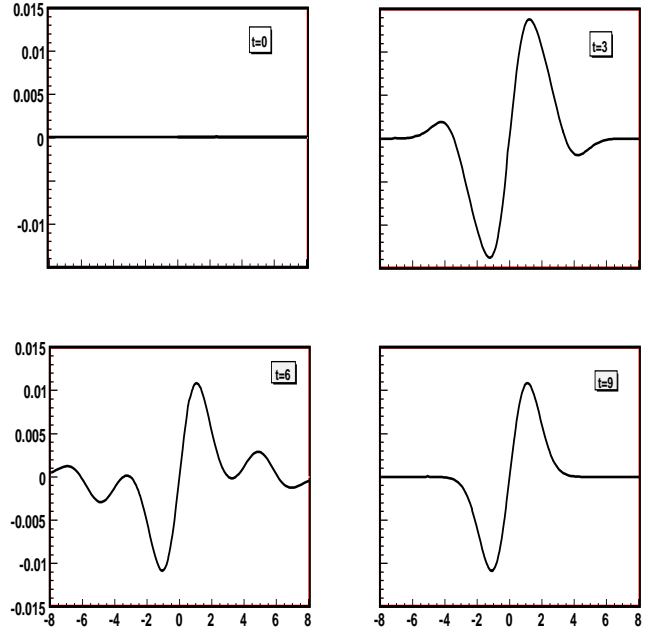,width=90mm,height=90mm}
\caption{Evolution of Minkowski spacetime with a non-trivial slicing
using an axi-symmetric code with a hyperbolic formulation. The plots
show the evolution of the $\Delta^\rho$ function at different times
along the $\rho$ axis.}
\label{fig:hy_delta} 
\end{figure}

Figures~\ref{fig:hy_alpha},~\ref{fig:hy_KA} and~\ref{fig:hy_delta}
show the evolution of the lapse function $\alpha$, the radial metric
$A$, and the variable $\Delta^\rho$ along the $\rho$ axis. We evolve
the system until $50M$ and all variables remain well behaved on the
axis.

%%%%%%%%%%%%%%%%%%%%%%%
%%%   BRILL WAVES   %%%
%%%%%%%%%%%%%%%%%%%%%%%

\subsection{Brill waves}

For our last example we have considered a non-trivial Brill wave
spacetime, which corresponds to strong non-linear gravitational waves
in vacuum. The construction of such a spacetime starts by considering
an axi-symmetric initial slice with a metric of the form
\begin{equation}
ds^2 = \Psi^4 \left[ e^{2q} \left( d\rho^2 + dz^2 \right)
+ \rho^2 d\phi^2 \right] \; ,
\label{eq:brillmetric}
\end{equation}
where both $q$ and $\Psi$ are functions of $(t,\rho,z)$ only.  In
order to solve for $\Psi$, we first impose the condition of time
symmetry, that is, $K_{ij}= 0$. This condition implies that the
momentum constraints~\eqref{eq:mom} are identically satisfied. We then
choose a specific form for the function $q$ and solve the Hamiltonian
constraint for $\Psi$, which for the metric~\eqref{eq:brillmetric}
becomes
\begin{equation}
\Delta_{\delta} \Psi + \frac{1}{4} \, \left( q_{,\rho\rho}
+ q_{,zz} \right) \Psi = 0 \; ,
\label{eq:brillham}
\end{equation}
with $\Delta_{\delta}$ the flat space Laplacian.  The function $q$ is
quasi-arbitrary, and must only satisfy the following boundary
conditions
\begin{eqnarray}
q \left|_{\rho=0} \right. &=& 0 \; ,\\
\partial^n_\rho q \left|_{\rho=0} \right. &=& 0 \qquad
\mbox{for odd $n$} \; , \\
q \left|_{r\rightarrow\infty} \right. &=& O \left( r^{-2} \right) \; .
\end{eqnarray}
Once a function $q$ has been chosen, all that is left for one to do is
to solve the elliptic equation~\eqref{eq:brillham} numerically.

\begin{figure}
\epsfig{file=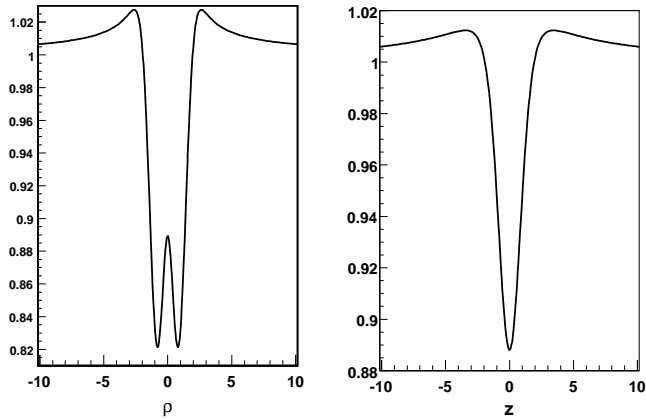,width=90mm}
\caption{Conformal factor $\Psi(\rho,z)$, along of the $\rho$ and $z$
axis respectively, for Brill initial data with a source function $q$
of the form~\eqref{eqn:holzq} and an amplitude of $a=2$.}
\label{fig:psi} 
\end{figure}

Different forms of the function $q$ have been used by different
authors~\cite{Eppley77,Holz93}.  Here we will consider the one
introduced by Holz and collaborators in~\cite{Holz93}, which has the
form
\begin{equation}
q = a \rho^2 e^{-(\rho^2 + z^2)} \; ,
\label{eqn:holzq}
\end{equation}
with $a$ a constant that determines the initial amplitude of the wave
(for small $a$ the waves disperse to infinity, while for large $a$
they can collapse to form a black hole).  Figure~\ref{fig:psi} shows
the value of $\Psi$ along the equator $z=0$ and axis $\rho=0$ obtained
by solving equation~\eqref{eq:brillham} numerically for an amplitude
of $a=2.0$ (small enough so that no black hole is formed, but large
enough so that we are far from the linear regime).

For the evolution of this initial data we have used a grid spacing of
$\Delta\rho=\Delta z=0.125$ and a Courant factor of $\Delta t / \Delta
\rho=0.2$, with the outer boundaries located at \mbox{$\rho=165,z=\pm
165$}. For the lapse evolution we use a 1+log slicing condition, which
corresponds to a Bona-Masso~\eqref{eq:BonaMasso} slicing with
$f(\alpha)=2/\alpha$.  Again, Kreiss-Oliger second order dissipation
has been added for stability.

\begin{figure}
\epsfig{file=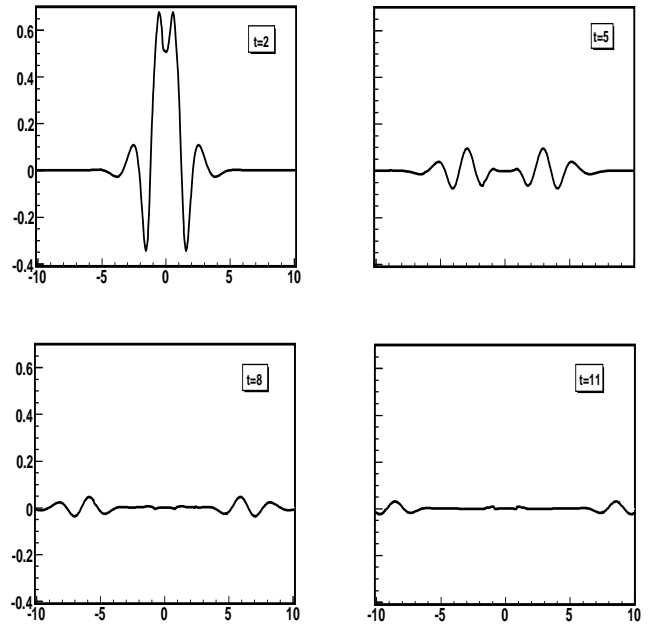,width=90mm,height=90mm}
\caption{Brill wave simulation. The plots show the evolution of the
extrinsic curvature component $K_{zz}$ at different times along the
$\rho$ axis.}
\label{fig:B_KB} 
\end{figure}

\begin{figure}
\epsfig{file=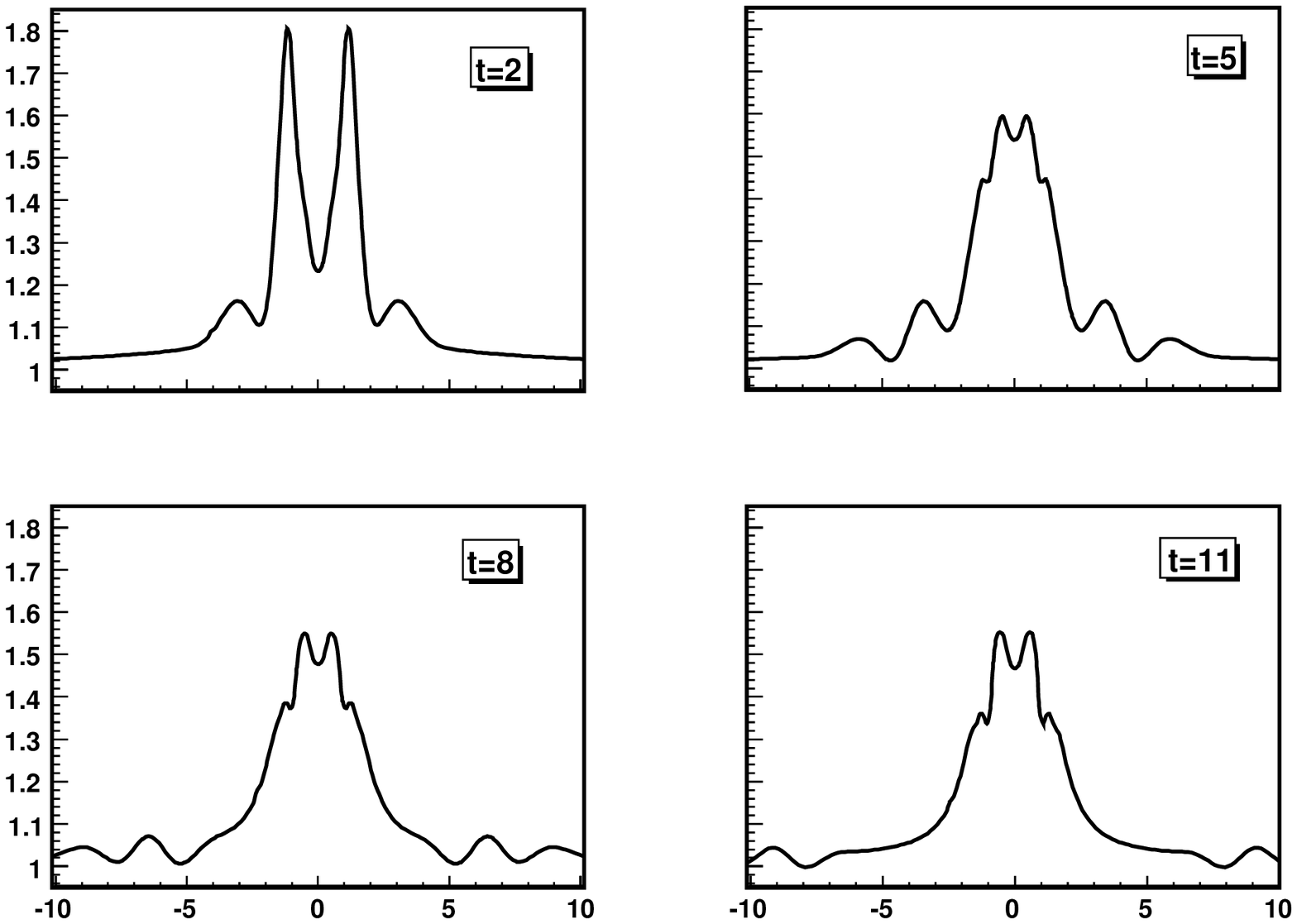,width=90mm,height=90mm}
\caption{Brill Wave simulation.  The plots show the evolution of the
metric function  $B$ at different times along the
$\rho$ axis.}
\label{fig:B_B} 
\end{figure}

\begin{figure}
\epsfig{file=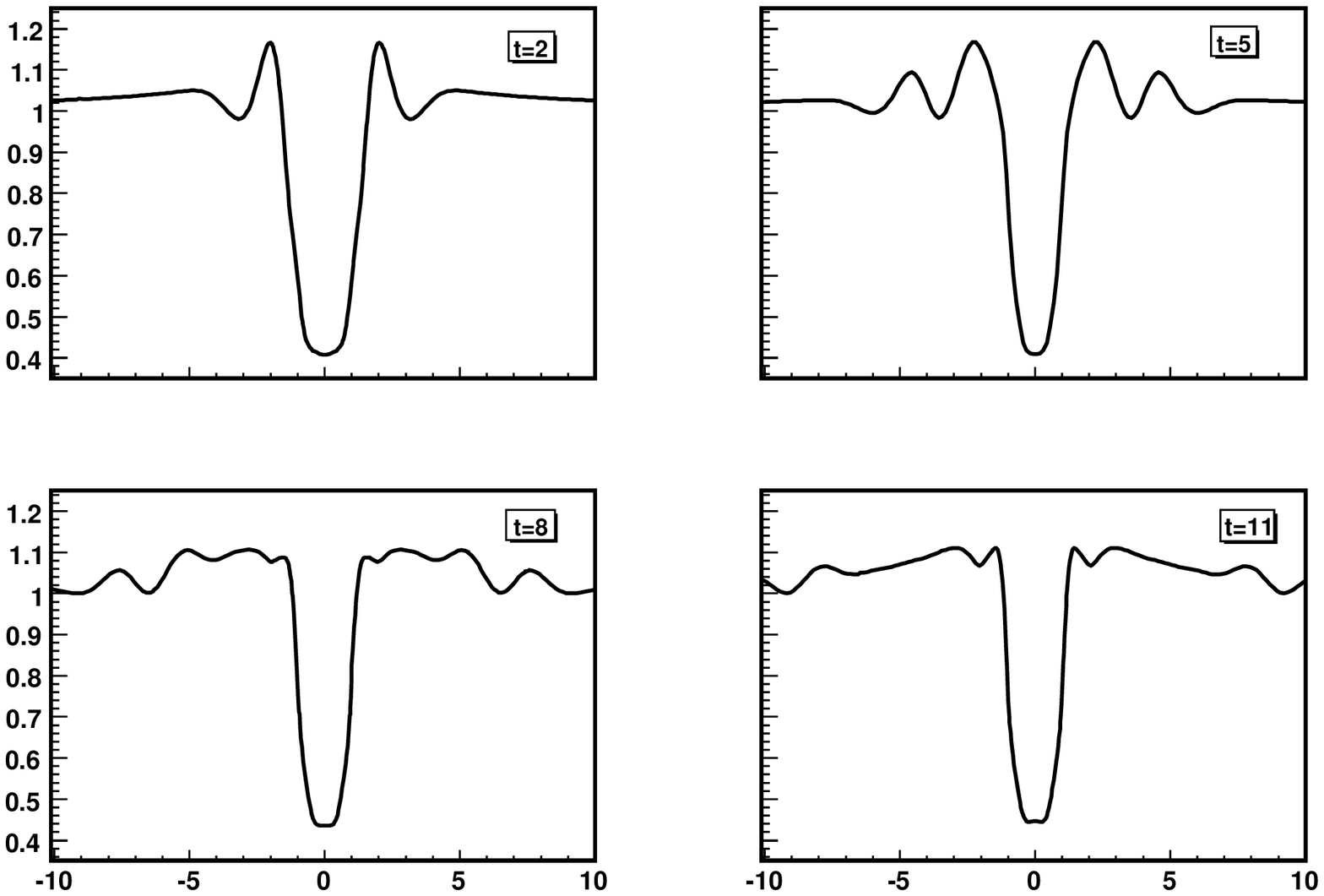,width=90mm,height=90mm}
\caption{Brill Wave simulation.  The plots show the evolution of the
metric function $T$ at different times along the $\rho$ axis.}
\label{fig:B_gammaz} 
\end{figure}

Figures~\ref{fig:B_KB}, \ref{fig:B_B} and~\ref{fig:B_gammaz} show the
evolution of the extrinsic curvature component $K_{zz}$, and the
metric components $B$ and $T$ along the $\rho$ axis.  Notice again
that for this simulation there is no problem at the axis of symmetry
in the evolution of the different geometric quantities.

%%%%%%%%%%%%%%%%%%%%%%
%%%   DISCUSSION   %%%
%%%%%%%%%%%%%%%%%%%%%%

\section{Discussion}
\label{sec:discussion}

We have presented a regularization procedure for the numerical
simulation of spacetimes with either spherical or axial symmetry
following an idea of Rinne and Stewart~\cite{Rinne:2005sk}. This
procedure enforces both the parity conditions and the conditions
arising from local flatness at the origin and the axis of symmetry.
We paid particular attention to the fact that our regularization
procedure is independent of the system of evolution equations chosen,
explicitly showing this in the case of the ADM formulation, as well as
a strongly hyperbolic formulation similar to that of Nagy, Ortiz and
Reula~\cite{Nagy:2004td} (slightly modified in order to have all the
dynamical variables well defined in curvilinear coordinates).

We have also described numerical codes that follow such regularization
procedure both in spherical and axial symmetry, and presented several
examples clearly showing that all dynamical variables remain regular
at the origin and axis of symmetry in each case (similar numerical
experiments using the regularized Z4 system
of~\cite{Rinne:2005sk,Rinne:2005df} have also been carried out by
Rinne and Stewart in~\cite{Rinne:2005df}). These results show that one
can construct well behaved numerical codes in both spherical and axial
symmetry that can allow the study of interesting astrophysical systems
with quite modest computer resources by using symmetry adapted
coordinate systems.

We conclude by mentioning that one can also construct regular codes
using specialized gauge choices which can even allow one to reduce the
number of independent components of the metric.  Nevertheless, our
interest here has been to find a regularization procedure that works
in the most general case, while leaving the choice of gauge completely
arbitrary.

%%%%%%%%%%%%%%%%%%%%%%%%%%%
%%%   ACKNOWLEDGMENTS   %%%
%%%%%%%%%%%%%%%%%%%%%%%%%%%

\acknowledgments

This work was supported in part by Direcci\'on General de Estudios de
Posgrado (DEGP), by CONACyT through grants 47201-F, 47209-F, and by
DGAPA-UNAM through grant IN113907.

%%%%%%%%%%%%%%%%%%%%
%%%   APPENDIX   %%%
%%%%%%%%%%%%%%%%%%%%

\appendix

%%%%%%%%%%%%%%%%%%%%%%
%%%   APPENDIX A   %%%
%%%%%%%%%%%%%%%%%%%%%%

\section{Evolution Equation in the  Spherical Case}
\label{app:spherical_equations}

In this appendix we give explicitly the evolution equations for the
hyperbolic system presented in the section~\ref{sec:hyperbolic} in
order to show that all equations are manifestly regular. We only
consider the case of spherical symetry, since in axi-symmetry the
final equations are just too long to write down here (they have been
calculated with Mathematica, and some of them are dozens of lines
long). For the spherical case, the resulting system of evolution
equations is, together with~\eqref{eq:S_variables}:

\begin{widetext}

\begin{eqnarray}
\partial_t K_H&=&
\frac{\alpha H^3}{2 r A^2 T^2} \,
\left( \Delta^r H^2 - F_r H-2 D_H \right) + \frac{\alpha}{2 A^2 T^2}
\Bigg[ \frac{{D_H}^2}{4} \left( {3 H^2} -r^2 J (A + 9 H) \right)
+ \frac{A r^4 {D_J}^2 T}{4} \nonumber \\
&+& H^2 J r^2 \left( J + 2 K_A K_H - 2 r F_r J \right)
+ \left( A^2 {D_H} T^2 - r J \left( r^2 H^3 J+r^2 A H J T
+ A^2  T^2 \right) \right) \, {\Delta^r} \nonumber \\
&+& r D_H J \left( 5 H^2 + 8 r^2 H J + 13 r^4 J^2 \right)
+ r^6 J^3 \left( -2 K_A {K_H} + 3 J \left( r F_r-5 \right) \right)
- A^2 {F_r}^2 T^2 \nonumber \\
&+& r^2 D_J \left( A F - H^2 J-5 r^5 J^3 \right)
+ 2 H^3 \left( J + 2 r F_r J + {K_A} \left( {K_T} - r^2{K_J} \right) \right)
- 4 A^2 T^2 \partial_r F_r \nonumber \\
&-& 2 r^4 H J^2 \left( 2 J \left( rF_r + 7 \right)
+ K_A \left( K_T - r^2 K_J \right) \right) + \frac{r^4 HD_J}{2} \,
\left( D_J H - 36 J^2 r - 5 D_J J r^2 \right) \nonumber \\
&+& \frac{1}{2} D_H D_J r^2 \left( 3 H^2 - 2 H J r^2 + 7 J^2 r^4 \right) T^2
+ A^3 T^2 \partial_r \Delta^r - A T^2 \partial_r{D_H} \Bigg] \, ,
\end{eqnarray}

\begin{eqnarray}
\partial_t K_J &=& \frac{\alpha H^5}{2 r A^2 T^2} \: \partial_r
\left( \frac{\Delta^r}{r} \right) - \frac{\alpha H^4}{2 r A^2 T^2} \:
\partial_r \left( \frac{F_r}{r} \right)
+ \frac{\alpha H^2 D_H}{8 r^2 A^2 T^2} \left( 4 F_r H-D_H \right)
+ \frac{\alpha H^2}{2 r  A^2  T^2} \nonumber \\
&\times& \left( -6 H D_J + 11 J D_H - 2 H J F_r + 3 J H^2 \Delta^r \right)
+ D_H^2 \left( -5 H J + \frac{3 J^2 r^2}{2} \right) \nonumber \\
&+& D_H \left( H^2 \left( 7 D_J - 2 J F_r \right)
+ 12 H J^2 r - 2 H J \left( D_J + J r^2 F_r \right) + 14 J^3 r^3
+ J^2 r^4 \left( 3 D_J + 2 J F_r \right) \right) \nonumber \\
&-& 4 H^3 K_A K_J + H J r^2 \left( -5 D_J^2 r^2
+ 4 J^2 \left( r F_r - 13 \right)
+ 4 J r \left( -6 D_J + K_A K_J r \right) \right) \nonumber \\
&+& 4 H^2 \left( J K_A \left( K_H + K_T \right) - 3 J r D_J
+ J^2 \left( 2 r^2 F_r^2 - 9 - \frac{D_J^2 r^2}{4} \right) \right)
- 2 A T^2 \partial_r D_J \nonumber \\
&-& J^2 r^4 \left( 8 J K_A \left( K_H + K_T \right)
+ 12 r J D_J - 3 D_J^2 r^2 + 4 J^2 \left( r F_r \left( r F_r + 1 \right)
+ 5 \right) \right) \nonumber \\
&+& 2 \left( A^2 D_J T^2 + J^2 r \left( -2 J r^2 + T \right) 
\left( H^2 + A T \right) \right) \Delta^r
+ 2 \left( 2 H^2 J^2 r^2 - J^4 r^6 \right) \partial_r F_r \nonumber \\
&+& 2 J \left( -H^2 \left( A + H \right) J r^2
+ A J^3 r^6 + A H^2 T \right) \, \partial_r \Delta^r \, ,
\end{eqnarray}

\begin{eqnarray}
\partial_t \Delta^r &=& \frac{\alpha}{A^3 T^2}
\Bigg[ 2 H^2 r \left( 6 J K_H + A K_J \right)
+ 12 J^3 r^5 K_H - 10 A J^2 r^5 K_J \nonumber \\
&+& A F_r T \left( -A K_J r^2 + K_H \left( A + 2 J r^2 \right) \right)
+ 2 D_H K_A T^2 + 2 D_J K_A r^2 T^2 - A^2 T \partial_r K_H \nonumber \\
&-& 2 H \left( J^2 \left( -4 K_T r^3 + 4 K_J r^5 \right)
+ A T \left( F_r K_J r^2 + \partial_r K_H \right) \right)
+ A \left( A + 2 J \right) r^2 T \partial_r K_J \Bigg] \, .
\end{eqnarray}

\end{widetext}

Considering the results of Section~\ref{sec:spherical}, we see that
the above equations are manifestly regular at the origin.

%%%%%%%%%%%%%%%%%%%%%%
%%%   REFERENCES   %%%
%%%%%%%%%%%%%%%%%%%%%%

\bibliographystyle{bibtex/apsrev}
\bibliography{bibtex/referencias}

%%%%%%%%%%%%%%%
%%%   END   %%%
%%%%%%%%%%%%%%%

\end{document}